\documentclass{article}
\usepackage{spconf,amsmath,graphicx}
\usepackage{hyperref}
\hypersetup{
}

\title{Vertebra-focused landmark detection for scoliosis assessment}

\name{Jingru Yi, 
        Pengxiang Wu, 
        Qiaoying Huang, 
        Hui Qu,
        Dimitris N. Metaxas
}
\address{Department of Computer Science, Rutgers University,
        Piscataway, NJ 08854, USA}

\begin{document}
%
\maketitle
\begin{abstract}
Adolescent idiopathic scoliosis (AIS) is a lifetime disease that arises in children. Accurate estimation of Cobb angles of the scoliosis is essential for clinicians to make diagnosis and treatment decisions. The Cobb angles are measured according to the vertebrae landmarks. Existing regression-based methods for the vertebra landmark detection typically suffer from large dense mapping parameters and inaccurate landmark localization. The segmentation-based methods tend to predict connected or corrupted vertebra masks. In this paper, we propose a novel vertebra-focused landmark detection method. Our model first localizes the vertebra centers, based on which it then traces the four corner landmarks of the vertebra through the learned corner offset. In this way, our method is able to keep the order of the landmarks. The comparison results demonstrate the merits of our method in both Cobb angle measurement and landmark detection on low-contrast and ambiguous X-ray images. Code is available at: \url{https://github.com/yijingru/Vertebra-Landmark-Detection}.
\end{abstract}
\section{Introduction}
\label{sec:intro}
\begin{figure}[t!]
  \centering
  \includegraphics[width=8.5cm]{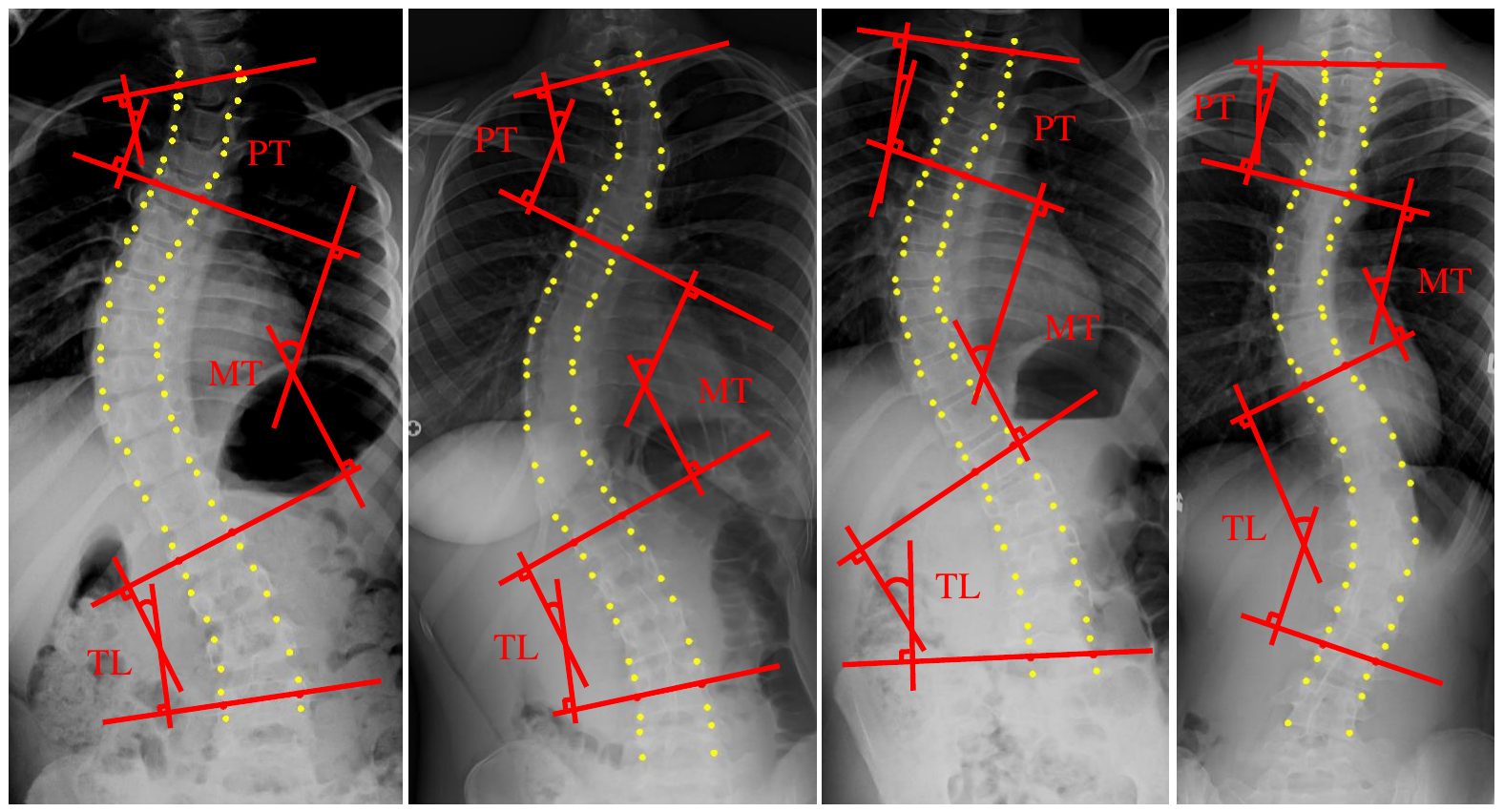}
\caption{Illustration of the anterior-posterior (AP) X-ray images. The ground-truth landmarks (68 points in total, 4 points per vertebra) are shown in yellow points. The coronal Cobb angles of proximal thoracic (PT), main thoracic (MT) and the thoracolumbar (TL) curves \cite{o2008radiographic,WANG2019101542} are measured according to the selected vertebrae. The red lines pass through the center lines of the selected vertebrae.
}
\label{fig:figure1}
\end{figure}

Adolescent idiopathic scoliosis (AIS) is a lateral deviation and axial rotation of the spine \cite{scholten1987analysis} that arises in children at or around puberty \cite{weinstein2008adolescent}. Early detection and bracing treatment of scoliosis would decrease the need for surgery \cite{anitha2012automatic}.  Cobb angle \cite{scholten1987analysis} is used as a gold standard by clinicians for scoliosis assessment and diagnosis. It is commonly measured based on the anterior-posterior (AP) radiography (X-ray) by selecting the most tilted vertebra at the top and bottom of the spine \cite{anitha2014automatic,anitha2012automatic}. Measurement of the Cobb angles is challenging due to the ambiguity and variability in the scoliosis AP X-ray images (Fig.~\ref{fig:figure1}). Generally, the clinicians manually measure the landmarks (the yellow points in Fig.~\ref{fig:figure1}) and choose the particular tilted vertebrae for the Cobb angle assessment. However, the measurement tends to be affected by the selection of the vertebrae and the bias of different observers.

Given that manual scoliosis assessment of Cobb angles in clinical practice is time-consuming and unreliable, there is a surge of interest in developing automatic methods for accurate spinal curvature estimation in spinal AP X-ray images. Traditional unsupervised methods such as filtering \cite{anitha2014automatic} and active contour \cite{anitha2012automatic} are parameter sensitive and typically involve complicated processing stages. To deal with the large anatomical variability and the low tissue contrast in X-ray images, supervised learning-based methods are developed. S$^2$VR \cite{sun2017direct} uses structured Support Vector Regression (SVR) to regress the landmarks and the Cobb angles directly based on the extracted hand-crafted features. BoostNet \cite{wu2017automatic} learns more robust spinal features by convolutional layers. These regression-based methods are able to exploit the global information of the image. However, the dense mapping between the regressed points and the latent features requires significant parameter and computational costs. Consequently, the input image ($\sim$2500$\times$1000) has to be downsampled to a very small resolution (e.g., 256$\times$128) to enable training and inference. Such an operation limits the performance of these methods due to the loss of fine details from the original high-resolution images. To handle this issue, another direction proposes to use convolutional layers to segment each vertebra for scoliosis assessment \cite{horng2019cobb,lu2018deepspine}. These methods are mainly based on U-net, and tend to be sensitive to the image qualities and difficult to separate the attached vertebrae.

\begin{figure*}[t!]
  \centering
  \includegraphics[width=0.98\textwidth]{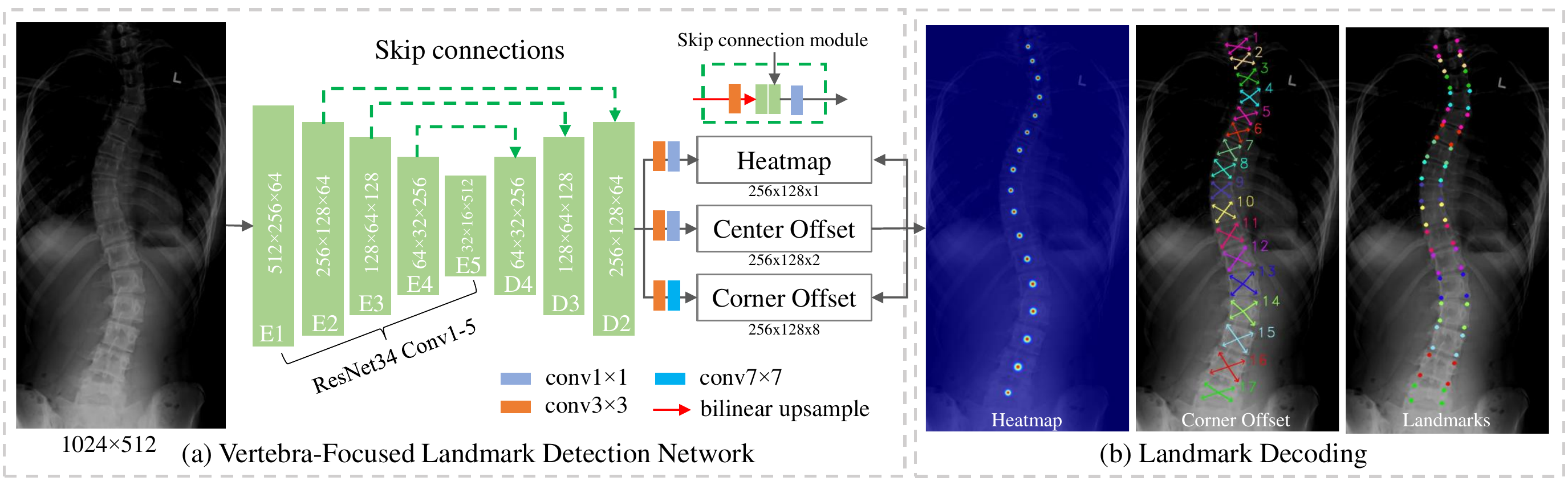}
\caption{(a)The framework of the vertebra-focused landmark detection network. The backbone of the network (i.e., conv1-5) is from ResNet34 \cite{he2016deep}. The sizes of the feature maps are presented as height$\times$width$\times$channels.
E and D represent encoder and decoder, respectively. The gray-scale input image is resized to 1024$\times$512. The skip connections combine the shallow and deep layers through the skip connection module. (b) Landmark decoding process. The vertebrae centers are extracted from the heatmap and the center offset. From the center of each vertebra, the four corner landmarks are traced using the corner offset.}
\label{fig:figure2}
\end{figure*}

Recently, keypoint-based methods have achieved remarkable performance in human pose joint localization \cite{newell2016stacked} and object detection \cite{law2018cornernet,yi2019multi,yi2019object}. Unlike the regression-based methods, the keypoint-based methods localize the points without dense mapping. Therefore, it simplifies the network and is able to consume the higher-resolution input image. In this paper, we propose a vertebra-focused landmark detection method based on keypoint detection. We make the network learn to differentiate different vertebrae by localizing the vertebra centers directly. After capturing the vertebrae, we regress the four corner landmarks of each vertebra using convolutional layers. In this way, we keep the order of the landmarks. Experimental results demonstrate the superiority of our method compared to the regression and segmentation based methods.

\section{Method}
As is shown in Fig.~\ref{fig:figure1}, the Cobb angles are determined by the locations of landmarks. The X-ray image we used contains 17 vertebrae from the thoracic and the lumbar spine. Each vertebra has 4 corner landmarks (top-left, top-right, bottom-left and bottom-right), and each image totally has 68 landmarks. The relative orders of landmarks are important for accurately localizing the tilted vertebrae. Considering this, we do not localize the 68 points directly from the output feature map since the model cannot guarantee that the detected points will stay at the right positions, especially when there are false positives, which would lead to incorrect landmark ordering. To address this issue, one strategy is to separate the landmarks into different groups, thus giving an output feature map with channel number $17 \times 4 = 68$. However, since each channel of output feature map has only one positive point, this strategy suffers from the class imbalance issue between the positive and negative points, which will hurt the model performance.

In this paper, we propose to first localize the 17 vertebrae by detecting their center points. One advantage of this approach is that the center points will not overlap. Therefore, the center points can be used to identify each vertebra without suffering from the touching problem in segmentation-based methods. After the vertebrae are localized, we then capture the 4 corner landmarks of each vertebra from its center point. In this way, we are able to keep the order of landmarks.

Our vertebra-focused landmark detection network is illustrated in Fig.~\ref{fig:figure2}. The inputs of the network are gray-scale images with size fixed to 1024$\times$512. First, we use ResNet34 \cite{he2016deep} conv1-5 to extract the high-level semantic features of the input image. Then we use the skip connections to combine the deep features with the shallow ones to exploit both high-level semantic information and low-level fine details, similar to \cite{ronneberger2015u,YI2019228}. At layer D2, we construct the heatmap, center offset and corner offset maps using convolutional layers for landmark localization.

\subsection{Heatmap of Center Points}
The keypoint heatmap is generally used in pose joint localization and object detection. For each point $k$, its ground-truth is an unnormalized 2D Gaussian disk (see Fig.~\ref{fig:figure2}b) which can be formulated as $exp({-\frac{x^2+y^2}{2\sigma^2}})$. The radius $\sigma$ is determined by the size of the vertebrae \cite{law2018cornernet}. We use the variant of the focal loss to optimize the parameters, the same as \cite{law2018cornernet,zhou2019bottom}:
\begin{equation}
    L_{hm} = -\frac{1}{N}\sum_{i}
\begin{cases}
    (1-p_i)^\alpha\log p_i& y_{i}=1\\
    (1-y_i)^\beta(p_i)^\alpha\log(1-p_i)              & \text{otherwise}
\end{cases},
\end{equation}
where $i$ indexes to each position of the feature map. $N$ is the total number of positions on the feature map, $p_i$ and $y_i$ refer to the prediction and ground-truth values, respectively. We set the parameters $\alpha=2$ and $\beta=4$ \cite{law2018cornernet} in this paper.

\subsection{Center Offset}
As can be seen from Fig.~\ref{fig:figure2}a, the output feature map of the network is downsized compared to the input images. This not only saves computational cost but also alleviates the imbalance problem between the positive and negative points due to the reduced output resolution. Consequently, a position $(x,y)$ on the input image is mapped to the location $(\lfloor\frac{x}{n}\rfloor,\lfloor\frac{y}{n}\rfloor)$ of the downsized feature map, where $n$ is the downsampling factor. After extracting the center points from the downsized feature map, we use the center offset to map the points back to the original input image. The center offset is defined as:
\begin{equation}
    (\frac{x}{n}-\lfloor\frac{x}{n}\rfloor,\frac{y}{n}-\lfloor\frac{y}{n}\rfloor).
\end{equation}
The center offsets at the center points are trained with L1 loss.

\subsection{Corner Offset}
When the center points of each vertebra are localized, we trace the 4 corner landmarks from the vertebra using corner offsets. The corner offsets are defined as vectors that start from the center and point to the vertebra corners (see Fig.~\ref{fig:figure2}b). The corner offset map has $4\times2$ channels. We use L1 loss to train the corner offsets at the center points.

\section{Experimental Details}

\subsection{Dataset}
We use training data (580 images) of the public AASCE MICCAI 2019 challenge as our dataset. All the images are the anterior-posterior X-ray images. Specifically, we use 60\% of the dataset for training (348 images), 20\% for validation (116 images), and 20\% for testing (116 images). Each image contains 17 vertebrae from the thoracic and lumbar spine. Each vertebra is located by 4 corner landmarks. The ground-truth landmarks (68 points per image) are provided by local clinicians. The Cobb angle is calculated using the algorithm provided by AASCE. The input images vary in sizes ($\sim$2500$\times$1000).

\subsection{Implementation}
We implement our method in PyTorch with NVIDIA K40 GPUs. The backbone network ResNet34 \cite{kingma2014adam} is pre-trained on ImageNet. Other weights of the network are initialized from a standard Gaussian distribution. We fix the input resolution of the images to $1024\times512$, which gives an output resolution of $256\times128$. To reduce overfitting, we adopt the standard data augmentation, including random expanding, cropping, contrast and brightness distortion. The network is optimized with Adam \cite{kingma2014adam} with an initial learning rate $2.5\times 10^{-4}$. We train the network for 100 epochs and stop when the validation loss does not decrease significantly.

\subsection{Evaluation Metrics}
Following the AASCE Challenge, we use the symmetric mean absolute percentage error (SMAPE) to evaluate the accuracy of the measured Cobb angles:
\begin{equation}
    \text{SMAPE}=\frac{1}{N}\sum_{j=1}^N\frac{\sum_{i=1}^3(|a_{ji}-b_{ji}|)}{\sum_{i=1}^3(a_{ji}+b_{ji})},
\end{equation}
where $i$ indexes the three Cobb angles in the area of proximal thoracic (PT), main thoracic (MT) and the thoracolumbar (TL), $j$ denotes the $j$-th image, and $N$ is the total number of testing images. The $a$ and $b$ refer to the estimated and the ground-truth Cobb angles, respectively. We also report the SMAPE for PT, MT and TL area individually which we represent as SMAPE$_{PT}$, SMAPE$_{MT}$ and SMAPE$_{TL}$.

We evaluate the accuracy of the landmarks by comparing the detected landmark locations to the ground-truth landmark locations. The averaged detection error is:
\begin{equation}
   \text{Error}_{dec} = \frac{1}{M}\sum_{i=1}^M||d_i-g_i||_2,
\end{equation}
where $d_i=(d_{x,i},d_{y,i})$ and $g_i=(g_{x,i},g_{y,i})$ are the detected and ground-truth landmark locations, respectively; $M$ is the total number of landmarks of the whole testing images.

\begin{figure*}[t!]
  \centering
  \includegraphics[width=0.98\textwidth]{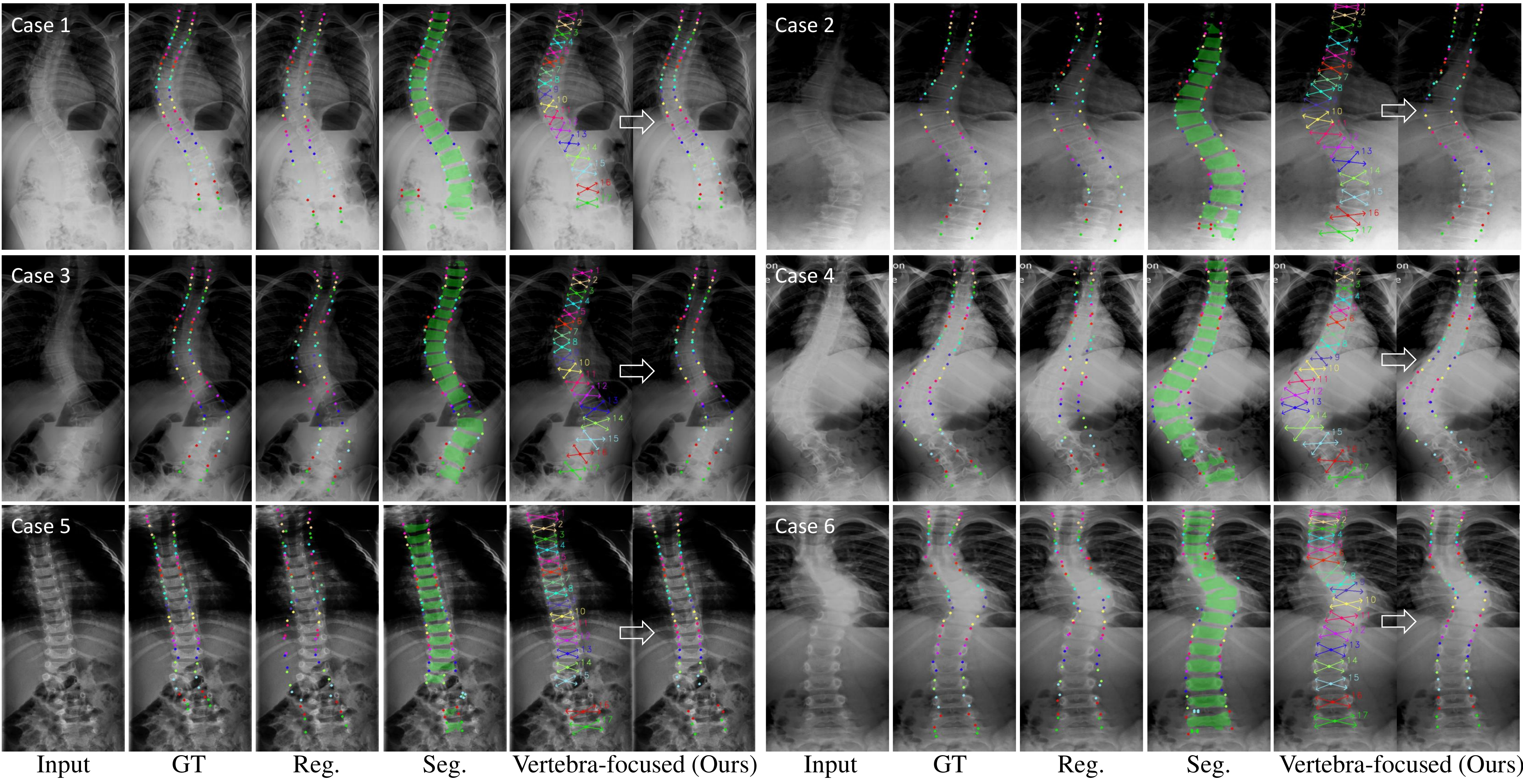}
\caption{Qualitative results of the landmark detection. GT refers to the ground-truth landmarks. For our method, we show both the prediction of the corner offsets and the landmarks. The big arrows indicate that the detected landmarks are from the four corner landmarks of the vertebra. The results of the segmentation-based method are overlaid with its predicted masks. The different color refers to different vertebra.}
\label{fig:figure3}
\end{figure*}

\begin{table*}[h!]
\begin{center}
\caption{Evaluation results of Cobb measurement and landmark detection. The frame per second (FPS) is measured on  K40.}
\begin{tabular}{lccccccc}
\hline
Method  & Input Resolution & SMAPE & SMAPE$_{PT}$ & SMAPE$_{MT}$ & SMAPE$_{TL}$ & Error$_{dec}$& FPS   \\ \hline
Regression-based  \cite{WANG2019101542} & $512\times256$ & 23.43 & 16.38  & 30.27  & 35.61  & 77.94          & \textbf{11.40} \\ \hline
Segmentation-based \cite{horng2019cobb} & $1024\times512$  & 16.48 & 9.71   & 25.97  & 33.01  & 74.07          & 2.38 \\ \hline
Ours  & $1024\times512$ & \textbf{10.81} & \textbf{6.26}   & \textbf{18.04}  & \textbf{23.42}  & \textbf{50.11}     & 5.65  \\ \hline
\end{tabular}
\label{tab:table1}
\end{center}
\end{table*}

\section{Results and Discussion}
We compare our method with the regression-based method \cite{WANG2019101542} and the segmentation-based method \cite{horng2019cobb}. The qualitative results and the quantitative results are shown in Fig.~\ref{fig:figure3} and Table~\ref{tab:table1}. Note that the regression-based method has a smaller input resolution because the parameters in the FC layers are too large and the GPU memory is limited. The landmarks of the segmentation-based method are decoded from the corner points of the minimum bounding rectangle of the vertebra segmentation mask. We use the same data augmentations and training skills for all these baseline methods.

As is shown in Fig.~\ref{fig:figure3}, the regression-based method performs well in capturing the orders of landmarks. This is owing to the separated channels of the FC layer. However, it fails to capture the landmark locations accurately. One reason would be that the small input resolution loses the morphology details of the vertebrae. In addition, the dataset we used is not large enough for the model to learn well as there are lots of parameters in the FC layers. Different from the regression-based method, the segmentation-based method captures the landmark locations better with the aid of segmentation masks. We show the overlayed vertebrae masks in Fig.~\ref{fig:figure3}. However, as can be seen from cases 1 and 6, the segmentation-based method fails to separate the connected regions. Moreover, for cases 2-4, the segmentation-based method tends to produce corrupted masks due to the ambiguity of the input images. Consequently, the false predictions disrupt the orders of detected landmarks and incur errors in landmark detection and Cobb angle calculation. This is also explained in Table \ref{tab:table1} that the segmentation-based method performs worse in the TL area of the spine as the vertebra typically gets more ambiguous in this part. In particular, in the TL area, the landmark error of the segmentation-based method is very close to that of the regression-based method.

Compared to the baseline methods, our vertebra-focused method achieves the best performance in both Cobb angle measurement (SMAPE) and the landmark detection (Error$_{dec}$), as shown in Table~\ref{tab:table1}. We illustrate both the corner offsets and the detected landmarks in Fig.~\ref{fig:figure3}. The corner offsets are colored arrows starting from the decoded center point of the vertebra. From cases 2,4,6, we can see that the vertebra-focused method is robust in localizing the vertebrae that have low-contrast in the original images. The reason would be that the model has the ability to identify the vertebra according to their global morphology features through center localization. We show a failure example in case 5, which suggests that the vertebra-focused network would skip the vertebra that has lower morphology property than the other vertebrae. However, such a failure does not affect the detection of the remaining vertebrae, indicating that the proposed method has better object reasoning ability.

\section{Conclusion}
\label{sec: conclusion}
In this paper, we propose a vertebra-focused landmark detection method that traces the corner landmarks of a vertebra from its center point. The strategy of predicting center heatmaps enables our model to identify different vertebrae and allows it to detect landmarks robustly from the low-contrast images and ambiguous boundaries. In contrast to the regression- and segmentation-based methods, our vertebra-focused method performs the best in landmark detection and Cobb measurement.


\end{document}